\documentclass[conference]{IEEEtran}
\usepackage[T1]{fontenc}
\usepackage[utf8]{inputenc}
\usepackage{cite}
\usepackage{amsmath,amssymb}
\usepackage{graphicx}
\usepackage{array,booktabs}
\usepackage{tikz}
\usetikzlibrary{arrows.meta,positioning,calc}
\usepackage[hidelinks]{hyperref}
\usepackage{url}
\title{Human-Aware Agentic Driving:\\Rethinking Driver State, Trust, and Intervention}
\author{\IEEEauthorblockN{Lu Feng}
\IEEEauthorblockA{Department of Computer Science\\University of Virginia\\Charlottesville, Virginia, USA}}
\begin{document}

\maketitle

\begin{abstract}
Agentic driving brings goal-directed reasoning, memory, tools, and adaptive interaction into automated vehicles. A vehicle can clarify a request, explain a maneuver, or prepare a driver to take control, then adapt its next action to the response. These capabilities make the vehicle an active participant in shaping the human behavior it observes. We examine how agentic interaction affects driver-state inference, trust and reliance, and human intervention in conditional automation. Established findings on attention, takeover prediction, trust-aware planning, and adaptive warnings provide useful foundations, but their applicability depends on how observations are elicited and how assistance changes the driving task. We distinguish what remains applicable, what changes under flexible interaction, and what remains unresolved across these three areas. The analysis identifies three research priorities: establishing the effects of interventions, jointly designing interaction and driving policies, and maintaining useful adaptation across repeated use and system changes. Progress should be measured through driver comprehension, appropriate reliance, and the ability to perform the required driving action in time.
\end{abstract}


\begin{figure*}[!t]
\centering
\begingroup
\definecolor{archInk}{HTML}{18334C}
\definecolor{archBlue}{HTML}{2F6BBD}
\definecolor{archTeal}{HTML}{008C95}
\definecolor{archOrange}{HTML}{C76A24}
\definecolor{archPaleBlue}{HTML}{EFF5FD}
\definecolor{archPaleTeal}{HTML}{EDF8F7}
\definecolor{archPaleOrange}{HTML}{FFF5EB}
\resizebox{\textwidth}{!}{%
\begin{tikzpicture}[x=1cm,y=1cm,>=Latex,font=\sffamily\small,text=archInk,
 panel/.style={rounded corners=7pt,line width=0.8pt},
 heading/.style={font=\sffamily\bfseries\large},
 item/.style={align=center,font=\sffamily\small},
 flow/.style={->,line width=1.15pt,rounded corners=6pt},
 flowlabel/.style={align=center,font=\sffamily\footnotesize,text width=1.55cm}]
\path[use as bounding box] (0,0.2) rectangle (17.55,5.8);
\filldraw[panel,fill=archPaleBlue,draw=archBlue!45] (0,1.1) rectangle (4.3,4.75);
\filldraw[panel,fill=archPaleTeal,draw=archTeal!45] (5.95,1.1) rectangle (11.65,4.75);
\filldraw[panel,fill=archPaleOrange,draw=archOrange!45] (13.3,1.1) rectangle (17.55,4.75);
\node[heading,text=archBlue] at (2.15,4.3) {Human};
\node[heading,text=archTeal] at (8.8,4.3) {Driving agent};
\node[heading,text=archOrange] at (15.425,4.3) {Vehicle + road};
\node[item,text width=3.75cm] at (2.15,3.52)
 {\textbf{Driver state}\\Attention and readiness};
\node[item,text width=3.75cm] at (2.15,2.64)
 {\textbf{Trust}\\Calibrated reliance};
\node[item,text width=3.75cm] at (2.15,1.76)
 {\textbf{Intervention}\\Preparation and takeover};
\node[item,text width=5.05cm] at (8.8,3.52)
 {\textbf{Understand}\\Scene and human state};
\node[item,text width=5.05cm] at (8.8,2.64)
 {\textbf{Reason}\\Goals, memory, and tools};
\node[item,text width=5.05cm] at (8.8,1.76)
 {\textbf{Act and adapt}\\Explain, assist, and coordinate};
\node[item,text width=3.7cm] at (15.425,3.3)
 {\textbf{Execute}\\Planning, safety checks, control};
\node[item,text width=3.7cm] at (15.425,2.05)
 {\textbf{Observe outcomes}\\Vehicle motion and traffic};
\draw[flow,archBlue] (4.3,3.25) -- (5.95,3.25);
\node[flowlabel,text=archBlue] at (5.125,3.72) {Goals + responses};
\draw[flow,archBlue] (5.95,2.25) -- (4.3,2.25);
\node[flowlabel,text=archBlue] at (5.125,1.78) {Dialogue + assistance};
\draw[flow,archOrange] (11.65,2.9) -- (13.3,2.9);
\node[flowlabel,text=archOrange] at (12.475,3.37) {Plans + requests};
\draw[flow,archTeal] (15.425,4.75) -- (15.425,5.3) -- (8.8,5.3) -- (8.8,4.75);
\node[font=\sffamily\small,text=archTeal] at (12.1,5.61) {Scene and action feedback};
\end{tikzpicture}%
}
\endgroup
\caption{Conceptual architecture of human-aware agentic driving. The human panel highlights driver state, trust, and intervention. Goals and observed responses inform the driving agent, which understands the scene and human state, reasons with memory and tools, and provides dialogue and assistance that can change attention, readiness, and reliance. Plans and requests pass to vehicle execution, while scene and action feedback inform the agent's next decisions.}
\label{fig:roadmap}
\end{figure*}
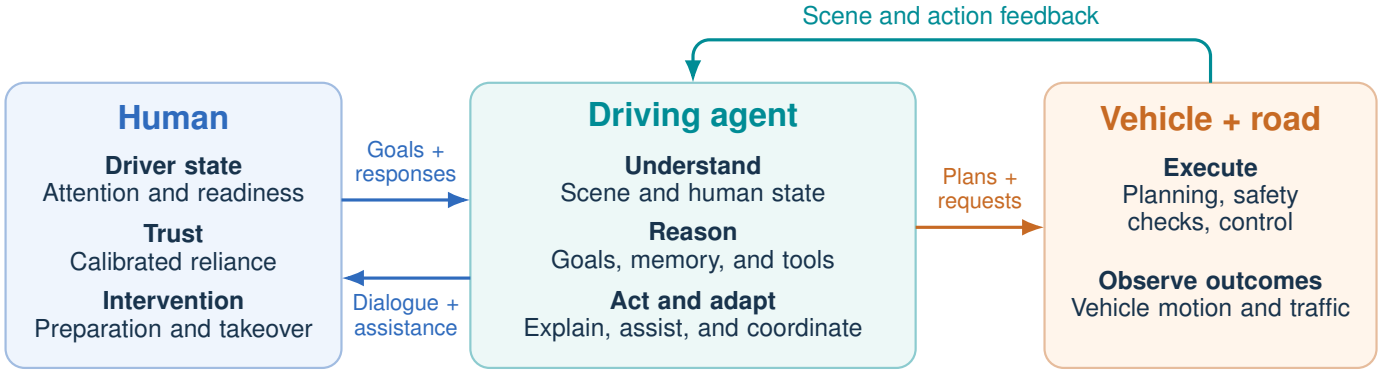

\section{Introduction}
Agentic approaches are attracting growing attention in automated-driving research, as language and multimodal models are incorporated into planning, assistance, and human--vehicle interaction \cite{Mao24,Yu25,WangMILD26,Nguyen25}. We use \emph{agentic driving} to describe goal-directed reasoning coupled with driving-related action and feedback. Its human implications are especially consequential in conditional automation, where a driver engaged in a non-driving activity may later need to resume control.

A vehicle approaching unfamiliar roadworks asks its driver to pause a film and look ahead. The driver glances at the road and replies, ``I'm ready.'' The glance may reflect understanding or simple compliance. The reply may express confidence in the vehicle, willingness to cooperate, or readiness to take control. A seemingly successful exchange therefore leaves several different inferences unresolved. The vehicle must interpret the response, judge how much time remains, and decide whether to continue preparing the person, request control transfer, or retain control.

The exchange illustrates how the vehicle helps produce the behavior it observes. Its questions can reveal information while changing attention, workload, and expectations. Interpreting the response therefore requires considering the interaction that preceded it.

Research on conditional automation already models several parts of this process. Task-sensitive attention models describe where people look \cite{Bae21}; multimodal predictors estimate takeover intention, time, and quality \cite{Pak21}; trust-aware planning represents a person's evolving confidence and its influence on decisions \cite{She19,She21,She22Trust,YuTrust24}; and context-sensitive warnings prepare drivers for intervention \cite{Pak22}. The issue is which of these models remain valid when the vehicle adapts its interaction with the driver.

Adaptive automation already changes attention and trust through warnings and feedback. Agentic systems can combine generated exchanges, response-dependent sequences, and memory across trips, extending that adaptation beyond the conditions studied with a particular warning or interface. The relevant issue is how the interaction policy changes the meaning of observations and the demands placed on the driver. A language model alone does not establish such a change.

Human--vehicle teaming frameworks organize cognitive capabilities \cite{WangPitts26}, and language-driven monitoring surveys organize systems and evaluation challenges \cite{Chandrayan26}. This paper contributes a structured synthesis of what earlier findings on driver state, trust, and intervention support, which assumptions become uncertain under flexible interaction, and which research priorities follow. The contribution lies in connecting these human-factors findings to agent design and evaluation. 
Figure~\ref{fig:roadmap} shows the human--agent--vehicle loop in which these decisions arise.
Table~\ref{tab:transfer} summarizes what stays, what changes, and what remains open.

\begin{table*}[!t]
\caption{Driver state, trust, and intervention in agentic driving: what stays, what changes, and what remains open.}
\label{tab:transfer}
\centering
\begingroup
\small
\setlength{\tabcolsep}{5pt}
\renewcommand{\arraystretch}{1.15}
\begin{tabular}{@{}>{\raggedright\arraybackslash}p{0.11\textwidth}>{\raggedright\arraybackslash}p{0.25\textwidth}>{\raggedright\arraybackslash}p{0.27\textwidth}>{\raggedright\arraybackslash}p{0.30\textwidth}@{}}
\toprule
\textbf{Theme} & \textbf{What stays} & \textbf{What changes} & \textbf{What remains open} \\
\midrule
\textbf{Driver state} &
Attention, comprehension, and control readiness remain distinct; multimodal prediction provides a foundation. &
Prompts and dialogue change the behavior being observed. Predictions depend on interaction history and candidate interventions. &
Separating revealed from induced behavior; predicting readiness under new interaction policies; reconciling remembered preferences with current capacity. \\
\midrule
\textbf{Trust} &
Appropriate reliance remains the goal; partially observable models connect evolving trust to decisions. &
Advice, conversation, and driving create multiple targets of trust. Success or failure in one capability may affect reliance on another. &
Determining when capability-specific beliefs are needed; measuring cross-capability effects; repairing expectations over repeated use. \\
\midrule
\textbf{Intervention} &
Content, timing, modality, and available response time shape preparation and takeover. &
The agent selects sequences of exchanges that consume attention and time while the vehicle continues to act. &
Choosing when to ask, assist, or remain silent; defining evidence for transfer; coordinating ambiguous responses with feasible vehicle actions. \\
\bottomrule
\end{tabular}
\endgroup
\end{table*}

\section{Agentic Driving and the Human's Role}
\label{sec:foundations}
\subsection{Agency in the driving and assistance loop}
Agentic driving couples goal interpretation, driving-related action, and revision through feedback. Actions include selecting a maneuver, consulting a tool, clarifying a request, and preparing a driver. Agent-Driver couples reasoning with tools and memory \cite{Mao24}; MILD connects cabin and road understanding to assistance; and SAFe-Copilot mediates between human and autonomous plans \cite{WangMILD26,Nguyen25}. Agency can therefore reside in planning, interaction, or their coordination, while a separate component retains execution authority. LanguageMPC links language-level decisions to low-level control \cite{Sha23}, while AgentsCoDriver combines communication, memory, and reflection for collaboration among vehicles \cite{Hu24b}. Their planning and coordination capabilities leave the effects on human readiness and reliance to be established.

Agency and automation level describe different aspects of a vehicle \cite{Yu25}. Flexible reasoning does not determine who performs the driving task or serves as fallback. Those responsibilities follow the operating arrangement. Here, the central setting is conditional automation, where the person can move from a non-driving activity to active control.

\subsection{Human roles and their implications}
The human participates in several roles. As a \emph{goal and preference provider}, the person specifies destinations, comfort preferences, or changes of plan. As an \emph{interaction partner}, the person supplies information, questions explanations, and negotiates assistance. As a \emph{driver available for control transfer}, the person may need to understand the scene and resume driving within a limited time. A driverless-service passenger can perform the first two roles without assuming the third. External road users present a further interaction setting, often without an established relationship with the agent.

These roles can overlap within a trip, but require different evidence. A preference indicates what the person wants; agreement indicates a response to a particular request; readiness concerns the capacity to perform the driving task. An agent that remembers a preferred driving style still needs current evidence of attention and capacity before requesting control transfer.

\section{Driver State in an Interactive System}
\label{sec:state}
\subsection{What stays: readiness has several components}
Agentic interaction preserves the distinction between looking at the road, understanding the scene, intending to intervene, and being able to control the vehicle. These components develop at different rates. Research on task interruption, attention recovery, and post-takeover stabilization shows why the first control input is an incomplete measure of readiness \cite{Eri17,Vog20,Gru24}. Felt readiness adds the driver's perspective without replacing behavioral evidence \cite{Sca21}. Urgency, secondary tasks, prior exposure, and warning modality remain relevant determinants \cite{Zha19}.

Existing models capture complementary parts of this process. Task-sensitive attention prediction, as in MEDIRL \cite{Bae21}, identifies where the driving task directs attention. Multimodal takeover models connect vehicle, physiological, and task observations to intention, response time, and control quality \cite{Pak21,Du20}. These targets and inputs remain useful when a driving agent is added. The distinction between preference and capacity also remains: personalized driving and uncertain-preference models describe desired behavior \cite{She22Style,ChenPref22}, while readiness concerns what the driver can do now. Remembering a preference for assertive progress does not establish present attention or capacity.

\subsection{What changes: interaction becomes part of the observation}
The agent can influence the evidence used to estimate readiness. At roadworks, a fixation on the blocked lane may be spontaneous, prompted, or part of a search initiated by a question. The gaze location is the same, but the inference depends on what the driver was asked and how they responded. A fixation predictor may remain accurate while its output becomes less informative about comprehension. Using it to support control transfer therefore requires evidence beyond gaze prediction.

The intervention also changes the response being predicted. A tactile cue, a brief explanation, and a request to identify the open lane impose different demands. A readiness estimate used to choose among them must account for the candidate action and the history that precedes it. The agent's memory of earlier answers, declined assistance, and current task engagement can matter as much as the latest sensor window.

These dependencies do not automatically invalidate existing models. If their inputs capture the relevant differences, refitting may suffice. If an unchanged predictor needs only adjusted probabilities or thresholds, recalibration is less extensive. Adding interaction history becomes justified when it explains response differences that the existing observations miss. The choice depends on which observations explain the response under the intended interaction.

\subsection{What remains open: when observed engagement supports action}
The central uncertainty is how reliably interaction-induced engagement predicts comprehension and control. Conversation can change subjective sleepiness and observable engagement on a test track \cite{Cockram26}; whether those changes improve takeover performance remains unresolved in the reported analysis. 

Evaluation needs to connect prompts and responses to independent comprehension measures and subsequent control. It should distinguish generalization to new drivers from generalization to a new interaction policy, and examine false predictions of readiness alongside average accuracy. Another unresolved issue is the cost of acquiring that evidence: asking a useful question can consume attention needed for the maneuver. A driver model should support the choice between observing, asking, and acting under uncertainty, while keeping persistent preferences separate from temporary capacity.

\section{Trust and Reliance Across Agent Capabilities}
\label{sec:trust}
\subsection{What stays: appropriate reliance is the objective}
Trust remains useful insofar as it helps explain and support appropriate reliance \cite{Lee04}. Reported trust, intention to delegate, actual use, and understanding of system limits are distinct constructs \cite{Wal23}. Readiness is distinct again: a person can willingly delegate while being unable to resume control, or distrust automation while remaining capable of intervention. A conversational acknowledgment cannot resolve all of these states.

Trust also remains dynamic and only partly observable. Event-related measurements and trust-aware planning provide ways to relate experience, uncertain attitudes, and decisions \cite{She19,She21,She22Trust,YuTrust24}. Transparency and workload already interact in adaptive interfaces \cite{Win20,Aka20}, and failures and takeovers affect trust over time \cite{Kra19}. These foundations apply to agentic driving; adaptive feedback and trust repair are established problems that acquire additional forms through conversation and memory.

\subsection{What changes: a shared interface spans capabilities}
Vehicles already offer capabilities that attract different levels of reliance. A conversational agent can present navigation advice, hazard explanations, and control-related assistance through a shared interface and remembered interaction history, making the boundaries between those capabilities less apparent. A general trust rating may obscure the resulting differences. Useful route suggestions could coexist with reluctance to delegate a difficult maneuver. Conversely, confidence earned through advice might influence driving reliance. The latter is a mechanism to investigate, rather than an established consequence of fluency.

The object of trust therefore needs to be identified in both measurement and planning. A shared latent trust state may still suffice if capability-conditioned observations and decision functions predict the relevant choices. Separate beliefs become useful when they explain consequential differences that recalibration or refitting within that representation misses. The belief-based formulation is a useful starting point; the fitted dynamics from a particular route-planning study do not establish performance across advisory and driving functions.

Generated explanations add another source of variation. Contrastive explanations connect a decision to alternatives and task-level reasons \cite{ChenExplain20}, but plausible wording can alter confidence without changing competence. Passenger negotiation similarly shows why satisfaction and agreement cannot stand in for warranted reliance: favorable ratings can accompany attempts to bypass a refusal to speed \cite{Stampf24}. An agent must communicate both what it can do and what it expects the person to do.

\subsection{What remains open: calibration across capabilities and time}
The unresolved issue is whether experience with one capability produces accurate or misplaced expectations about another. Separately varying advisory and driving performance, under comparable presentation, would help identify that influence. Subsequent delegation, monitoring, and responses to limitations provide stronger evidence about reliance than favorable ratings alone. Repeated observations are needed to determine whether a shared trust model remains adequate.

Failure and recovery make the distinction particularly important. An incorrect explanation, forgotten preference, and failed maneuver affect different expectations. Successful conversation after a driving failure could restore acceptance before driving competence improves. Repair should therefore be evaluated through later capability-specific choices as well as immediate reactions. Persistent memory must preserve consequential failures and corrections while updating stale expectations. What remains uncertain is which information supports justified recovery, how long cross-capability effects persist, and when personalization reinforces an inaccurate model of the vehicle.

\section{Communication, Preparation, and Control}
\label{sec:intervention}
\subsection{What stays: preparation must fit the driving task}
An intervention is useful when it helps the person understand and perform the required action within the available time. Content, timing, and modality remain central. Context-sensitive advisories demonstrate the value of designing preparation around the driver's task \cite{Pak22}; preliminary haptic and startle investigations examine additional perceptual and physiological aspects \cite{Pak18,Pak20}. Their different study scales support different levels of confidence. Improved subjective ratings of takeover information also need not imply improved objective situation awareness \cite{Col21}.

Preparation must be evaluated through the control transition. Available time constrains takeover outcomes \cite{Lia26}, and real-traffic observations show responses unfolding as sequences of actions \cite{Pip23}. Human-in-the-loop synthesis, predictive monitoring, and constrained planning provide ways to connect allocation and uncertain predictions to available vehicle actions \cite{Feng15,Feng16,Don25,She24Shield,She24ACP}. The need for such a connection persists when communication becomes more flexible.

\subsection{What changes: a message becomes a sequence of decisions}
An agent can ask a follow-up question, interpret the answer, and revise its next intervention. At roadworks, an advisory to look ahead might be followed by a question about the open lane. That exchange changes preparation time, cognitive demands, and the evidence available for transfer. The benefit and timing of a tested fixed advisory cannot simply be assigned to the resulting dialogue.

The interaction policy must represent waiting, ambiguous answers, and the option to stop asking. This policy extension need not change the human model if elapsed time and response quality are already represented. Otherwise, additional observations or state variables may be needed. Systems that connect cabin and road information to assistance or arbitrate between human and autonomous plans address parts of this interface \cite{WangMILD26,Nguyen25}; the message and the vehicle's behavior while awaiting a response must be coordinated.

Language also creates commitments. ``I will yield,'' ``please look ahead,'' and ``you can continue reading'' imply different vehicle actions and human responsibilities. Diagnostics and structured explanations connect statements to modeled requirements and alternatives \cite{FengDiag16,FengCounter18,ChenExplain20}; reasoning--trajectory monitoring addresses consistency with planned motion \cite{TianYu26}. Generated wording adds the possibility that a paraphrase changes implied certainty or permission. System consistency, human comprehension, and subsequent reliance consequently require separate evidence.

\subsection{What remains open: when dialogue improves the transition}
Evidence for adaptive intervention needs to connect the selected messages to the resulting transition. Profile-dependent warning generation establishes that assistance can be varied \cite{Xu25}; persuasive assistance and conversational studies provide evidence about secondary-task behavior, gaze, or engagement \cite{Xiang25,Cockram26}. The remaining step is to establish how those effects contribute to the complete transition to control.

Comparisons should isolate content, modality, timing, and personalization while tracking comprehension, workload, response delays, and subsequent control. They must also include silence, misunderstanding, and premature takeover, since those responses determine what the vehicle can do next. The unresolved design problem is how much evidence to seek before acting, and which vehicle actions remain feasible while it is sought. A useful dialogue policy would improve preparation without consuming the margin needed to complete or avoid transfer.

\section{Research Priorities for Human-Aware Agentic Driving}
\label{sec:roadmap}
The gaps in Sections~\ref{sec:state}--\ref{sec:intervention} require advances that connect human modeling to agent design. Three priorities deserve particular attention: evidence about intervention effects, policies that coordinate interaction with vehicle action, and adaptation across repeated use. Together they address the short-term response to a message, the driving decision made during an exchange, and the relationship that develops across trips.

\subsection{Data that support intervention decisions}
An agent needs evidence about which action helps a person in a given situation, beyond correlations between observed behavior and performance. A shared experimental resource should record the road and cabin context, agent actions, response timing, and subsequent human and vehicle outcomes. It should preserve the policy that selected each intervention, including when the agent chose silence. Otherwise, an apparent benefit of prompting may reflect that prompts were issued in easier situations or to more responsive drivers.

Randomized comparisons within suitable simulator conditions can estimate effects of message content and timing; observational data require explicit assumptions about how interventions were assigned. Evaluation splits should hold out interaction policies as well as participants. Attention and takeover predictors \cite{Bae21,Pak21} would then provide baselines for testing whether intervention history improves prediction and whether those gains support better action selection. The deliverable is evidence linking an intervention to its consequences, with uncertainty and limits of applicability, rather than a benchmark scored only on response prediction.

\subsection{Joint policies for interaction and driving}
Interaction selection and vehicle planning need a common account of time, human response, and remaining actions. A promising architecture would evaluate a proposed exchange alongside the vehicle behavior required while it unfolds. It would specify what evidence the exchange seeks, how long the agent can wait, and which action follows an ambiguous or absent response. A question is worthwhile only if its expected contribution to the driving decision exceeds the attention and time it consumes.

Trust-aware planning and explicit control-allocation methods provide starting points \cite{She21,She22Trust,YuTrust24,Feng15,Feng16}. Extending them requires uncertain human responses to affect both communication choices and feasible vehicle actions. Safety constraints on the latter \cite{She24Shield,She24ACP} must hold while the exchange is in progress, not only after an answer arrives. The critical comparison is joint selection against independently designed dialogue and driving policies under the same sensing and time budget. Relevant outcomes include unnecessary interruptions, missed preparation opportunities, and control failures following confident but incorrect readiness estimates.

\subsection{Adaptation that survives repeated use and system change}
Persistent memory makes personalization a continuing inference problem. Driver habits can change, stored preferences can become stale, and software updates can alter vehicle competence. The agent must decide which history remains relevant while the person is also revising expectations of the system. A profile that improves prediction during familiar trips may mislead both participants after a capability change.

Longitudinal evaluations should therefore include changes in driving capability, corrections to remembered preferences, and recovery after failures. They should compare persistent memory with recent-history and no-memory baselines, measuring whether adaptation improves current decisions and whether drivers recognize changed limits. Human-model updates, memory revisions, and software versions need to be traceable so that a deterioration in reliance or readiness can be attributed to the relevant change. This would establish when personalization remains useful, when old information should be discarded, and how the vehicle should communicate a change in what the person can expect.

\section{Conclusion}
Research on driver state, trust, and intervention provides a foundation for human-aware agentic driving. Flexible dialogue and persistent memory extend familiar adaptive-automation problems by changing how human observations are elicited, how capabilities are presented, and how preparation unfolds. Earlier principles remain useful, while fitted predictions and intervention benefits require evidence under the intended interaction conditions.

This paper identifies three priorities: establishing intervention effects, coordinating interaction and driving policies, and maintaining useful adaptation across repeated use and system change. Their value should be judged by whether drivers understand the situation, rely appropriately on the available capabilities, and can perform the required action in time. These outcomes connect advances in agent capabilities to the human responsibilities on which conditional automation depends.

\end{document}